\documentclass[twocolumn,showpacs,pra]{revtex4}
\usepackage{delarray}
\usepackage{amsmath, amssymb}
\usepackage{bm}
\usepackage{graphicx}
\usepackage{placeins}
\usepackage{color}
\usepackage[figuresright]{rotating}
\usepackage{float} 
\usepackage{lipsum}

\begin{document}
\title{Assessment and Enhancement of {SAR} non-coherent Change Detection Techniques Following Oil Spills}

\author{Cihan Bay\i nd\i r$^1$\footnote{Email: cihan.bayindir@isikun.edu.tr},  J. David Frost$^2$, Christopher F. Barnes$^3$}

\address{$^1$Assistant Professor, Department of Civil Engineering, I\c{s}\i k University, \c{S}ile, \.{I}stanbul, 34980, Turkey\,\\
\email{cihan.bayindir@isikun.edu.tr}}

\address{$^2$Professor, Department of Civil and Environmental Engineering,\\
\email{david.frost@ce.gatech.edu}}

\address{$^3$Associate Professor, Department of Electrical and Computer Engineering, Georgia Institute of Technology, Atlanta, Georgia, 30332, USA}

\begin{abstract}
Oil spill is one of the most dangerous catastrophes that threaten the oceans. Therefore detecting and monitoring the oils spill by means of remote sensing techniques which provide a large scale assessment is of critical importance to predict, prevent and clean the oil contamination. In this study the detection of the oil spill using synthetic aperture radar (SAR) imagery is considered.  Detection of the oil spill is performed using change detection algorithms between imagery acquired at different times. The specific algorithms used are the correlation coefficient change statistic and the intensity ratio change statistic algorithms. Therefore these algorithms and the probabilistic selection of the threshold criteria is reviewed and discussed. A recently offered change detection method which depends on the idea of generating two different final change maps of two images in a sequence, is used. First final change map is obtained by cumulatively adding the sequences of change maps in such a manner that common change areas are excluded and uncommon change areas are included. The second final change map is obtained by comparing the first and the last images in the temporal sequence. This method requires at least three images to be employed and can be generalized to longer temporal image sequences. The purpose of this approach is to provide a double check mechanism to the conventional approach and thus to reduce the probability of false alarm and enhance the change detection. The algorithms mentioned are applied to a 2010 Gulf of Mexico oil spill imagery together with the method described. It is shown that intensity ratio change statistic is a better tool for identification of the changes due to the oil spill compared to the correlation coefficient change statistic. It is also shown that two final change map method can reduce the probability of false alarm. The data used in this study is acquired by the Japanese Aerospace Agency's Advanced Land Observing Satellite (ALOS) through Alaska SAR Facility (ASF) at the University of Alaska, Fairbanks, AK.

\pacs{92.20.Ny, 93.30.Mj}
\end{abstract}
\maketitle

\section{Introduction}

SAR imaging is extensively used for monitoring various types of natural hazards and disasters. These include but are not limited to earthquakes and faults, landslides, volcanic activity, flooding, fires, hurricanes, tsunamis and oil spills \cite{Jackson}. Satellite systems became invaluable tools for early warning and post disasters assessment studies since they provide a large scale view of the disaster area. This is critically important to identify the effected areas by the disaster. Compared to some small scale monitoring systems, they can also provide information about the extent of the disaster in the inhabited areas. Various types of sensors, imaging modalities and image formation techniques are used in satellite systems which are used for monitoring the disasters. A review of the satellite sensors and associated image processing techniques used for monitoring the natural hazards and disasters can be seen in \cite{Joyce2009}.

SAR imagery is also used for monitoring the oceans. The first satellite used for acquiring imagery from the ocean environment and the first civilian satellite with a SAR sensor was the SEASAT satellite. It was launched by USA in 1978. After that, many satellites such as ERS-1, ERS-2, RADARSAT-1, RADARSAT-2, ENVISAT and ALOS were placed on their orbits for monitoring both the ocean and land surface. Measurements of  ocean surface waves and spectra, wave refraction and breaking, wind speed and direction, upwelling, sea ice, ocean currents and current gradients, oceanic internal waves, marine atmospheric boundary layers, underwater topography and detection of ships and wakes are just a few examples where SAR imaging is used for oceanographic studies \cite{Jackson}. SAR imagery is also used to detect oil spills and surfactants \cite{Jackson}.

Oil on the ocean surface can be released from ships, result from seepage from sea bottom or can be spilled from offshore platforms after collapses, fires or even under normal operating conditions. Bragg scattering is the dominant mechanism of the electromagnetic wave scattering from the ocean surface. Short gravity-capillary waves on the ocean surface are the most responsible components of the ocean wave spectra for the radar back scatter from ocean surface \cite{Jackson}. Mineral oil on the ocean surface damps these gravity-capillary waves. Therefore the radar back scatter reduces and oil appears as dark patches in the SAR imagery \cite{Jackson}.

Oil spill on the ocean surface can be treated as a two layer flow in terms of hydrodynamics. Realistic ocean wave models such as given in \cite{Jackson, bay2009, Karjadi2012} can be extended for two layer flow including the capillary effects which are characterized by surface tension and curvature of the ocean surface to model the oil spill on the ocean surface. Some similar studies can be seen in \cite{Jackson} where generally approximate ocean models are used. While these studies can provide a better understanding of the physics which governs the scattering of microwaves from the ocean surface they may not be the best way to understand the larger dynamics of oil spill on the ocean surface since they do not provide a global assessment.

In this paper, the detection of the oil spill by SAR imagery is discussed. The change detection algorithms are used to compare and detect the changes due to oil spill. The specific algorithms used for this purpose are the correlation coefficient change statistics and the intensity ratio change statistics. A review of these algorithms is presented. These algorithms are applied to data and an assessment of the two algorithms is provided. Additionally, a recently offered methodology which can reduce the probability of false alarms associated with detection of the changes is applied. This method depends on the idea of generating two change maps of any two images in a three image sequence. This methodology is applied to 2010 Gulf of Mexico oil spill data which is acquired by PALSAR of the ALOS satellite.

\section{Change Detection Methodology}
Comparisons between imagery acquired at different times can be made by some mathematical techniques known as change detection (CD) algorithms. They are used to extract information about the changes occurred in a scene. These changes can be in the magnitude, location and the direction of the scene. CD algorithms are widely used in digital image and video processing to detect temporal changes in one or multiple scenes \cite{bay2013}. They have the capability to detect very subtle changes both in the intensity and the phase values which may not be seen by the human eye. For ideal change detection, the imagery should be acquired by the same or similar sensor and using the same resolution, viewing angle and geometry, spectral bands, time of the day and season \cite{bay2013, bay2014, bay2015}. 

Change detection mapping and analysis can be applied to various types of imagery obtained by different sensing systems such as radar, optical, multispectral or lidar, just to name a few. CD can be applied in various disciplines including but not limited to remote sensing, surveillance, medical imaging, civil infrastructure and underwater sensing. SAR imagery is widely used for both civilian and military reconnaissance. Earthquake, flood, tsunami, hurricane damage assessment, monitoring oil spills, extreme snow and ice conditions, fires, deforestation, illegal ship traffic, structural damages, military constructions and operations are a few examples to mention where the airborne and satellite SAR imagery are widely used.

Currently there are many change detection algorithms available for use. It is possible to create a classification matrix or tensor of change detection algorithms. One possibility is to make a classification based on the dimensions of the imagery, as 2D or 3D change detection algorithms. 
Generally, depending on the phase values used or not, the CD algorithms are classified as non-coherent or coherent change detection algorithms. 

Non-coherent change detection identifies changes in the mean back scattered power of the scene by comparing sample estimates of the mean back scatter power taken from the repeat pass image pair however coherent change detection also accounts for the changes in the phases. In this paper only non-coherent change detection is considered. Two very widely used change detection algorithms, the correlation coefficient change statistic and the intensity ratio change statistic algorithms are used to make an assessment of the changed areas due to oil spill. A recently offered methodology which first appeared in \cite{bay2013} and depends on generating double change map in a temporal sequence with three images is used to reduce the false alarm probabilities. A review of these algorithms and the double change map methodology is provided in the next section.

\subsection{Review of the Correlation Coefficient Change Statistic}
The correlation coefficient between image pairs is defined as 
 \begin{equation}
   \widehat{\gamma}_{_{total}}=\widehat{\gamma}_{_{SNR}}\widehat{\gamma}_{_{base}}\widehat{\gamma}_{_{scene}}\widehat{\gamma}_{_{vol}}\widehat{\gamma}_{_{proc}}
  \label{CD1}
\end{equation}
where $\widehat{\gamma}_{_{SNR}}$, $\widehat{\gamma}_{_{base}}$, $\widehat{\gamma}_{_{scene}}$, $\widehat{\gamma}_{_{vol}}$ and $\widehat{\gamma}_{_{proc}}$ are the decorrelations due to relative back scatter signal to receiver noise ratio, due to mismatch in the acquisition geometries between first and repeat-pass data collections, in the scene,  due to scattering from a volume when a nonzero baseline is used in the acquisition and image processing, respectively. Discussions about the partial contributions of these phenomena to the correlation coefficient can be seen in \cite{Rignot1993}. Generally $\widehat{\gamma}_{_{SNR}}\widehat{\gamma}_{_{base}}\widehat{\gamma}_{_{vol}}\widehat{\gamma}_{_{proc}} \approx 1$ so that $\widehat{\gamma}_{_{scene}}$ is the indicator of the correlation between the image pair. From now on for the sake of brevity $\widehat{\gamma}$ is used in this paper instead of $\widehat{\gamma}_{_{scene}}$.

For image pairs with intensities $f(x,y)$ and $g(x,y)$ the correlation coefficient, $ \widehat{\gamma}$,  is defined as
 \begin{equation}
   \widehat{\gamma}(x,y)=\frac{\left|\sum_{i=1}^N {f_i g_i} \right|}{ \sqrt{\sum_{i=1}^N \left|f_i  \right|^2  \sum_{i=1}^N \left|g_i  \right|^2}}
  \label{CD2}
\end{equation}
where $N$ is the number of pixels in the neighborhood of the pixel $(x,y)$ and generally taken as $N=3 \times 3$ or $N=5 \times 5$ \cite{Preiss2006}. Cauchy-Schwarz inequality guarantees that $ \widehat{\gamma} \in [0,1]$. $\widehat{\gamma}=1$ refers to no change whereas $\widehat{\gamma}=0$ refers to a very significant change in the scene. When the coherent change detection (CCD) methods are adopted, the phase values of the received signals becomes important. For a valid correlation coefficient for CCD, the complex conjugate of either the first or the second image should be taken in the Eq.~\ref{CD2} \cite{bay2013}. 

A threshold is used to vote for if there is change or not between image pair. The selection of threshold exhibits a tradeoff between the probabilities of detection ($P_{d}$) and false alarm ($P_{fa}$) therefore related probability density functions (pdf) need to be used. The pdf of the correlation coefficient statistic is derived in \cite{Preiss2006, Touzi1988} as
\begin{equation}
   p(\widehat{\gamma} \vert  \gamma,N) =2(N-1)(1-\gamma^2)^N \widehat{\gamma} (1-\widehat{\gamma}^2)^{N-2} \ _2F_1(N,N;1;\gamma^2 \widehat{\gamma}^2)    \label{CD4}
\end{equation}
where $\gamma$ is the underlying scene coherence. It is given by
\begin{equation}
   \gamma=E \{(m_r)(n_r) \}  \ \  m_r=f/ \sigma_f  \ \   n_r=g/ \sigma_g
  \label{CD3}
\end{equation}
with $E \{I_f \}=\sigma^2_f$, $E \{I_g \}=\sigma^2_g$. $m_r$ and $n_r$ are the unit speckle noise components and $E\{ \}$ is the expected value operator. $_1F_2$ is the Gauss' hypergeometric function which is defined as 

 \begin{equation}
\begin{split}
  _1F_2(a, b; c; z)&= F(b,a; c; z)= \\
	& \frac{\Gamma(c)}{\Gamma(a)\Gamma(b)} \sum_{l=0}^\infty \frac{\Gamma(a+l) \Gamma(b+l)}{\Gamma(c+l)}\frac{z^l}{l!}
	\end{split}
   \label{CD15}
\end{equation}
where $\Gamma$ is the gamma function \cite{bay2013}. The plot of this pdf can be seen in the Figure~\ref{corrpdf1} below.

\begin{figure}[htb!]
\begin{center}
   \includegraphics[width=3.4in]{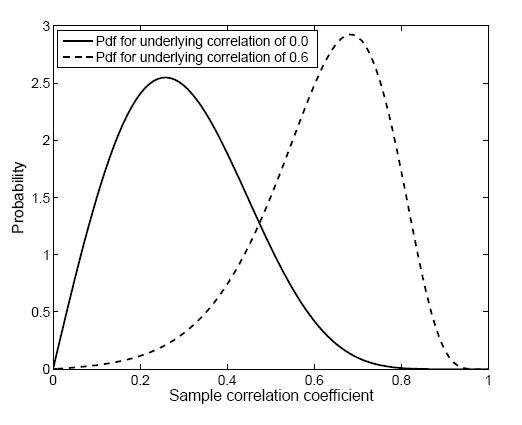}
  \end{center}
\caption{\small Pdf of correlation coefficient change statistics for an underlying scene coherence of $0.0$ and $0.6$ for $N=9$.}
  \label{corrpdf1}
\end{figure}

\begin{figure}[h]
\begin{center}
   \includegraphics[width=3.4in]{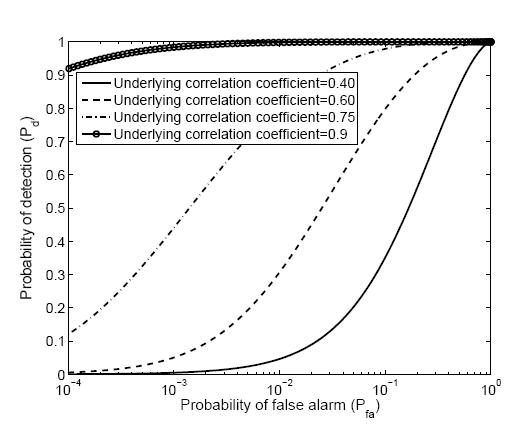}
  \end{center}
\caption[ROC curve for correlation coefficient change statistic obtained with underlying unchanged correlation coefficients of $0.40, 0.60, 0.75, 0.90$ and changed correlation coefficient of $0.0$ for $N=9$]
  {ROC curve for correlation coefficient change statistic obtained with underlying unchanged correlation coefficients of $0.40, 0.60, 0.75, 0.90$ and changed correlation coefficient of $0.0$ for $N=9$.}%
  \label{corrpdf2}%
\end{figure}

This pdf is tested with experimental data and since it agrees well with data it is widely used in change detection studies such as in \cite{Touzi1988, Preiss2006}. Based on this pdf, the $P_{fa}$ and $P_{d}$ can be calculated as \cite{Preiss2006}
\begin{equation}
\begin{split}
 & P_{fa} = \int_0^T p(\widehat{\gamma} \vert  \gamma=\gamma_{unchanged})d\widehat{\gamma} \\
 & P_{d} = \int_0^T p(\widehat{\gamma} \vert  \gamma=\gamma_{changed}) d\widehat{\gamma} 
\end{split}
   \label{CD5}
\end{equation}
which is equal to 
\begin{equation}
 \begin{split}
  P & = \frac{2(N-1)(1-\gamma^2)^N}{\Gamma(N)\Gamma(N-1)}.  \\
  & .\sum_{k=0}^{N-2} [ {N-2 \choose k} (-1)^{N-2-k} \sum_{l=0}^{\infty} \left[\frac{\Gamma(N+l)}{\Gamma(l+1)}  \right]^2 \gamma^{2l} \\
	&  {}{}.\frac{T^{2N+2l-2-2k}}{2N+2l-2-2k} ]    
  \end{split} 
   \label{CD6}
\end{equation}
which gives the $P_d$ and $P_{fa}$ as
\begin{equation}
   P= 
   \begin{cases}
P_d, & \text{for } \gamma=\gamma_{changed}, \\
P_{fa}, & \text{for }  \gamma=\gamma_{unchanged}. 
\end{cases}
   \label{CD7}
\end{equation}
For $\gamma_{changed}=0$, $P=P_d$ becomes \cite{Preiss2006}
\begin{equation}
  P =P_d = 2(N-1)^2 \sum_{k=0}^{N-2} {N-2 \choose k} (-1)^{N-2-k} \frac{T^{2N-2-2k}}{2N-2-2k}      
   \label{CD8}
\end{equation}
Threshold to be used for the change detection by the correlation coefficient change statistic algorithm can be decided by making use of $P_d$ vs. $P_{fa}$ graphs. In the Figure~\ref{corrpdf2} $P_d$ vs. $P_{fa}$, (or sometimes referred to as Receiver Operating Characteristics (ROC)) curves for $\gamma_{changed}=0$ and $\gamma_{unchanged}=0.40,0.60,0.75,0.90$ for $N=9$ are given. $P_d=0.8$ and $P_{fa}=0.1$ can be achieved for $\gamma_{unchanged}=0.60$ as it can be seen in the Figure~\ref{corrpdf2}. 

\begin{figure}[H]%
  \centering%
    \includegraphics[width=3.4in]{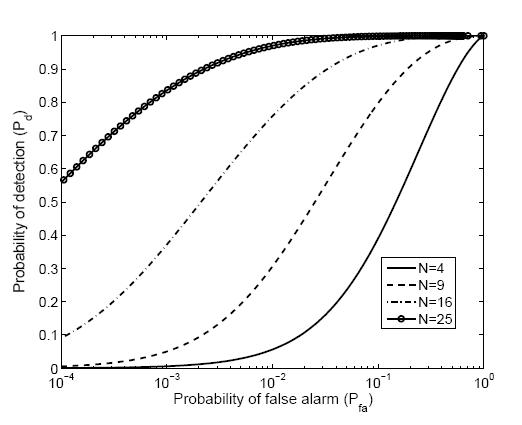}%
  \caption[ROC curve for correlation coefficient change statistic obtained with underlying unchanged correlation coefficients of $0.6$ for $N=4,9,16,25$]
  {ROC curve for correlation coefficient change statistic obtained with underlying unchanged correlation coefficients of $0.6$ for $N=4,9,16,25$.}%
  \label{corrpdf3}%
 \end{figure}

Generally a $P_{fa}$ which is on the order of $10^{-2}$ is used in the literature \cite{Preiss2006}.It is possible to reduce the $P_{fa}$ significantly, at least one order of magnitude, by making the sampling window larger \cite{Preiss2006}. In Figure~\ref{corrpdf3}, $P_d$ vs. $P_{fa}$ (ROC) curves for $\gamma_{changed}=0$ and $\gamma_{unchanged}=0.60$ for $N=4,9,16,25$ is presented. As it can be seen on the figure, increasing $N$ from 16 to 25 provides an order of magnitude reduction in the false alarm rate \cite{Preiss2006}.  However the window size should be commensurate with the disturbance size in the scene in order to prevent contributions from both changed and unchanged pixels therefore $N$ should be limited \cite{Preiss2006}.

\subsection{Review of the Intensity Ratio Change Statistic}
\begin{figure}[H]%
  \centering%
    \includegraphics[width=3.4in]{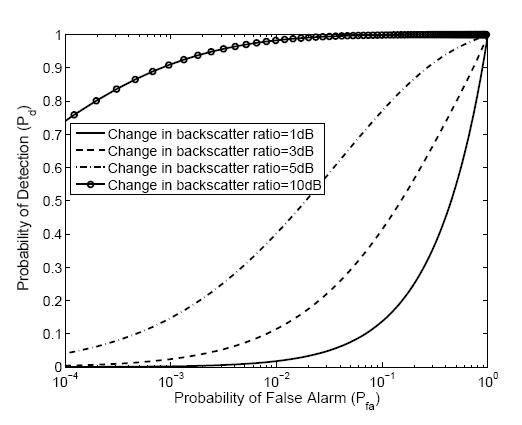}%
  \caption[ROC curve for the intensity ratio change statistic obtained with $R_0=0$ dB and $R_1=1$ dB, 3 dB, 5 dB, 10 dB for $N=9$]
  {ROC curve for the intensity ratio change statistic obtained with $R_0=0$ dB and $R_1=1$ dB, 3 dB, 5 dB, 10 dB for $N=9$.}%
  \label{ratiopdf2}%
\end{figure}

\begin{figure}[H]%
  \centering%
    \includegraphics[width=3.4in]{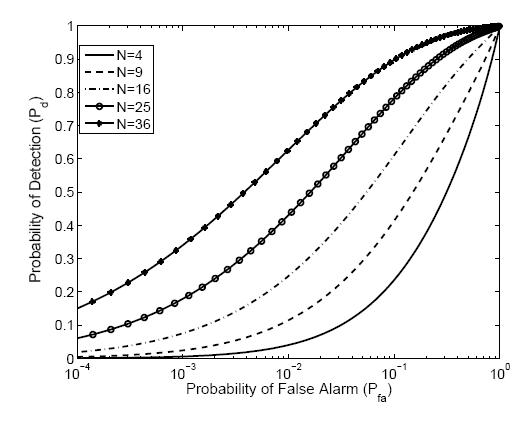}%
  \caption[ROC curve for the intensity ratio change statistic obtained with $R_0=0$ dB and $R_1=3$ dB for $N=4, 9, 16, 25, 36$]
  {ROC curve for the intensity ratio change statistic obtained with $R_0=0$ dB and $R_1=3$ dB for $N=4, 9, 16, 25, 36$.}%
  \label{ratiopdf3}%
 \end{figure}

The second change detection algorithm used in this paper is the intensity ratio change statistic first proposed by \cite{Touzi1988}. In this section a review of this algorithm which was first presented by \cite{Preiss2006} is given. The intensities of the image pair can be computed by 
\begin{equation}
   I_{f}=\frac{1}{N}\sum_{i=1}^N \vert f_i \vert^2  \ \ \ \ \ \ I_{g}=\frac{1}{N}\sum_{i=1}^N \vert g_i \vert^2 
  \label{CD9}
\end{equation}
The pdf of these intensities follow a gamma distribution which is verified using experimental data \cite{Rignot1993, Preiss2006, Touzi1988 }. The ratio change statistic is given by
\begin{equation}
   \widehat{R}=\frac{I_f}{I_g}
   \label{CD10}
\end{equation}
which has various forms offered in the literature. A successful version of this detector was first proposed in \cite{Touzi1988} and is given by
\begin{equation}
   \widehat{r}= 
   \begin{cases}
\widehat{R}, & \text{if } \widehat{R} \leq 1, \\
{\widehat{R}}^{-1}, & \text{if }  \widehat{R} > 1. 
\end{cases}
   \label{CD11}
\end{equation}
The $\widehat{r}$ rests in $[0,1]$ and a simple threshold can be applied to detect the changed and unchanged areas \cite{Preiss2006}. The pdf of this variable can be written as \cite{Preiss2006,Touzi1988},
\begin{equation}
   p(\widehat{r} \vert  R)=\frac{\Gamma(2N)}{\Gamma^2(N)} \left(\frac{R^N}{(\widehat{r}+R)^{2N}}+ \frac{R^{-N}}{(\widehat{r}+R^{-1})^{2N}} \right) \widehat{r}^{N-1}
   \label{CD12}
\end{equation}
where $E$ denotes the expected value and $R=E \{I_f \}/ E \{I_g \} $. The plot of this pdf can be seen in the Figure~\ref{ratiopdf1} below.
\begin{figure}[H]%
  \centering%
    \includegraphics[width=3.4in]{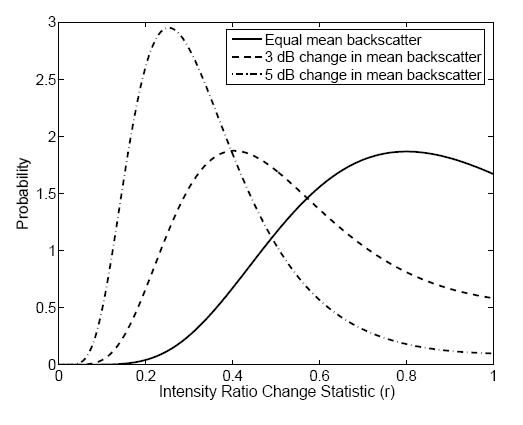}%
  \caption[Pdf of intensity ratio change statistics for an unchanged scene ($0$ dB) and scenes with $3$ dB and $5$ dB change in the back scatter for $N=9$]
  {Pdf of intensity ratio change statistics for an unchanged scene ($0$ dB) and scenes with $3$ dB and $5$ dB change in the back scatter for $N=9$.}%
  \label{ratiopdf1}%
 \end{figure}
This pdf is tested against various experimental test imagery and it successfully represents the data \cite{Preiss2006, Touzi1988}. Based on this pdf it is possible to calculate the $P_{fa}$ as \cite{Preiss2006}
\begin{equation}
 \begin{split}
  P_{fa}& = \int_0^T p(\widehat{r} \vert  R=R_0) d\widehat{r} \\
         &= \frac{\Gamma(2N)}{\Gamma^2(N)} \int_0^T  (\frac{R_0^N}{(\widehat{r}+R_0)^{2N}}+ \frac{R_0^{-N}}{(\widehat{r}+R_0^{-1})^{2N}} )\widehat{r}^{N-1} d\widehat{r} 
  \end{split} 
   \label{CD13}
\end{equation}
where $R_0=E \{I_f \}/ E \{I_g \}$ with $E \{I_f \}$ and $E \{I_g \}$ are the mean back scattered power in the unchanged regions of the image \cite{Preiss2006}. The $P_{d}$ becomes
\begin{equation}
 \begin{split}
  P_{d}& = \int_0^T p(\widehat{r} \vert  R=R_1) d\widehat{r} \\
         &= \frac{\Gamma(2N)}{\Gamma^2(N)} \int_0^T  (\frac{R_1^N}{(\widehat{r}+R_1)^{2N}}+ \frac{R_1^{-N}}{(\widehat{r}+R_1^{-1})^{2N}} )\widehat{r}^{N-1} d\widehat{r} 
  \end{split} 
   \label{CD14}
\end{equation}
where $R_1=E \{I_f \}/ E \{I_g \}$ with $E \{I_f \}$ and $E \{I_g \}$ are the mean back scattered power in the changed regions of the scene \cite{Preiss2006}. Analytical evaluations of these integrals are given in \cite{Preiss2006}. The evaluation of $P_{fa}$ and $P_{d}$ leads to the ROC curves given the Figure~\ref{ratiopdf2} and Figure~\ref{ratiopdf3}.

It is useful to note that for $P_d=0.7$ and $N=9$, intensity ratio change statistics with $R_1=3$ dB and correlation coefficient change statistic with $\gamma_{unchanged}=0.60$ shows a quite similar behavior with both yields to the value of $P_{fa}= 0.07$. Additionally it can be observed from the Figure~\ref{ratiopdf2} above that, the back scatter ratio change ($R_1$) of 3 dB and $P_d=0.7$ yields $P_{fa}\approx 0.35$ which is an unacceptably high value \cite{Preiss2006}. One possibility to overcome this high false alarm rate is to increase the $N$.
As it can be seen in Figure~\ref{ratiopdf3}, increasing $N$ from 9 to 25 reduces the $P_{fa}$ from 0.35 to 0.06, for  $P_d=0.7$ and $R_1=3$ dB. However since the disturbance size and the window size should be commensurate in order to prevent contributions from both changed and unchanged pixels, it is not possible to increase $N$ unboundedly \cite{Preiss2006}.

\subsection{Double Change Map Method for Improving the Probabilities of False Alarm} 
In this section a recently offered change detection methodology which relies on the idea of generating two final change maps in order to reduce the probability of false alarm is used. Although it is applied for the CD algorithms in this paper, it can be extended to CCD algorithms as well. Using a temporal sequence of $n$ images of the same target area, two final change maps are generated. First change map is obtained by cumulatively adding the successive change maps and the second change map is obtained by applying the CD algorithm to the first and the last image in the temporal sequence \cite{bay2013}.

When performing the cumulative addition common changed areas between the change map sequence is excluded to prevent accounting for a change first appears and then disappears. However uncommon changed areas between the change map sequence is included in an absolute manner. By adopting this approach two different change maps comparing first and the last image can be generated. The changes present in both of the change maps are critical since they are more likely to indicate a detected change, not a false alarm. The changes which are not present in both of the change maps are more likely to be false detections. Joint change maps are generated by taking the intersection of the two final change maps. Since more than one change maps are generated, sampling and the accuracy of the change maps increases. This reduces the false alarm rates for a given probability of detection. This method can provide a double-check mechanism for majority of the CD or CCD algorithms used in the literature. This approach is summarized by the flowchart in \cite{bay2013}. Due to available data, this method is applied to temporal sequences with three images \cite{bay2013}. 

\section{Results and Discussion}

The oil spill images used in this paper are acquired in 2010 at three different locations in the Gulf of Mexico. A review of the 2010 Gulf of Mexico oil spill can be seen in \cite{bay2013}. The locations of the acquired imagery and the location of the 2010  Gulf of Mexico oil spill can be seen in the Figure~\ref{GeneralView1} below.

\begin{figure}[h]
\begin{center}
   \includegraphics[width=3.4in]{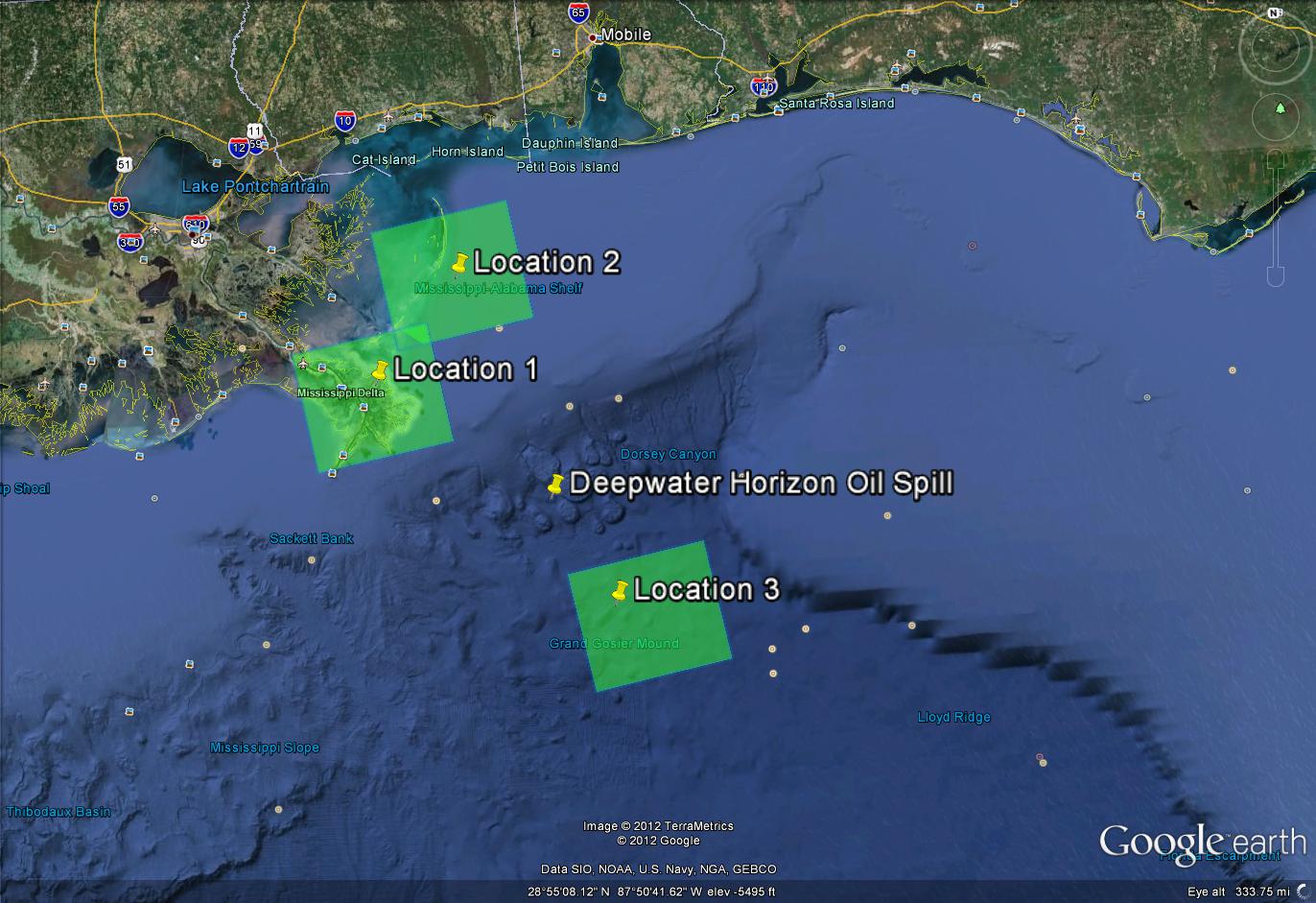}
\end{center}
\caption{Locations of satellite data for Gulf of Mexico Oil Spill.}
\label{GeneralView1}
\end{figure}

At each location three different images acquired at different times are used and therefore a total of nine images are analyzed in this paper. The scene identification numbers, acquisition dates, path and frame numbers and polarizations of these imagery are tabulated in \cite{bay2013}.

\begin{figure}[h!]
\begin{center}
   \includegraphics[width=3.4in]{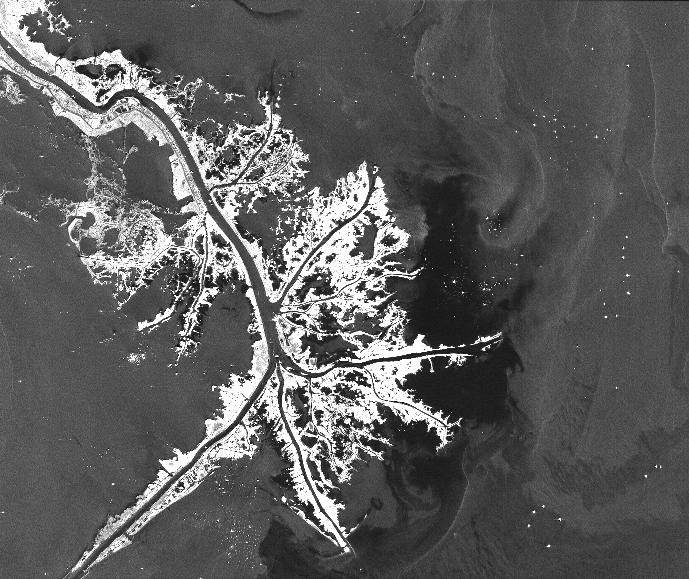}
\end{center}
\caption{Image ALPSRP229470570-Acquisition Date 05/16/2010}
\label{image_ALPSRP229470570}
\end{figure}

\begin{figure}[h!]
\begin{center}
   \includegraphics[width=3.4in]{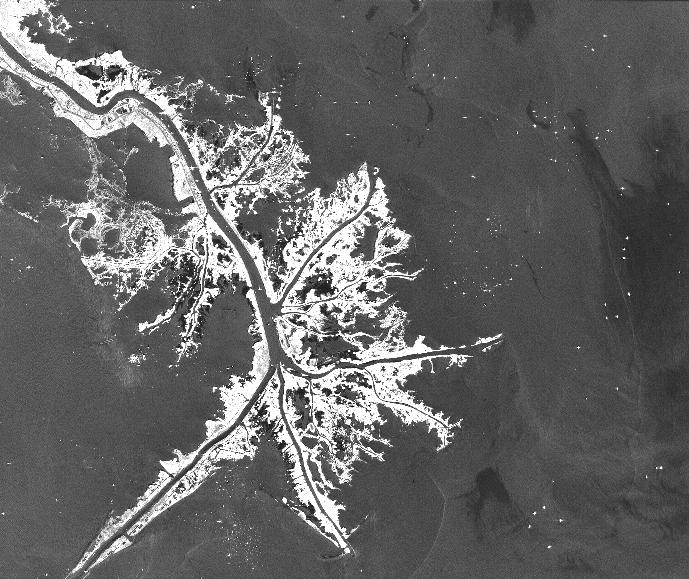}
\end{center}
\caption{Image ALPSRP236180570-Acquisition Date 07/01/2010.}
\label{image_ALPSRP236180570}
\end{figure}

\begin{figure}[h!]
\begin{center}
   \includegraphics[width=3.4in]{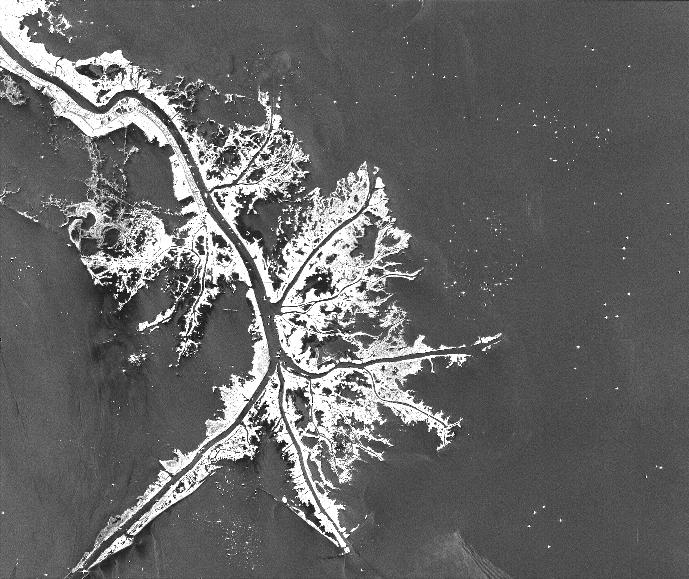}
\end{center}
\caption{Image ALPSRP242890570-Acquisition Date 08/16/2010}
\label{image_ALPSRP242890570}
\end{figure}
These imagery are acquired by the PALSAR of the ALOS satellite. Further discussion about ALOS and its sensors can be seen in \cite{Rosenqvist2004, Hamazaki}. Figures~\ref{image_ALPSRP229470570},~\ref{image_ALPSRP236180570} and~\ref{image_ALPSRP242890570} which can be seen above and below are the co-registered images acquired at location 1. The oil spill patches in the vicinity of the peninsula can be seen with the naked eye as it can be realized from the figures.

\begin{figure}[h!]
\begin{center}
   \includegraphics[width=3.4in]{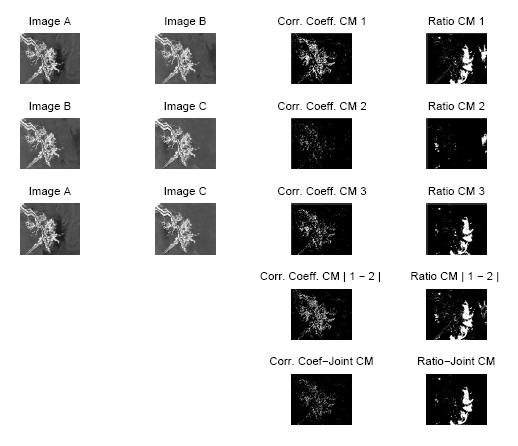}
\end{center}
\caption{Change maps of oil spill imagery collected on location 1.}
\label{CMGulf1}
\end{figure}

\begin{figure}[h!]
\begin{center}
   \includegraphics[width=3.4in]{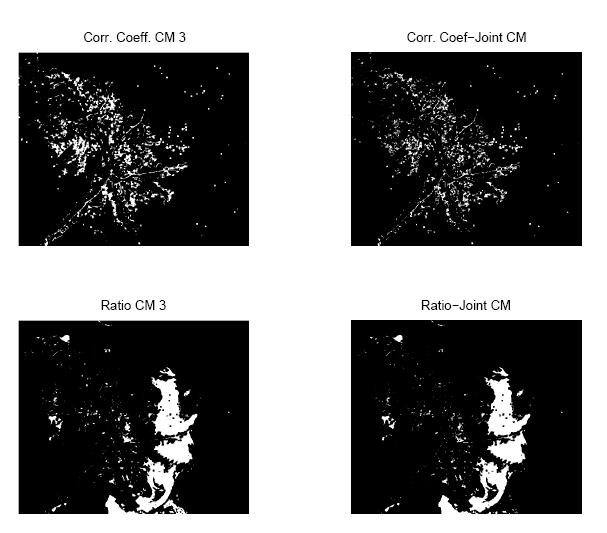}
\end{center}
\caption{Change maps of oil spill imagery collected on location 1.}
\label{CMGulf12}
\end{figure}

 For the correlation coefficient change statistics algorithm, the representative changed and unchanged areas in the images lead to $\gamma_{changed} \approx 0.45$ and  $\gamma_{unchanged}=0.92$, respectively. This result shows that the correlation between images are very high, even in the significantly changed regions. Therefore the change detection becomes harder. Additionally the correlation coefficient change statistic is more sensitive to the changes in the areas which are represented by bright pixels compared to the changes in the areas represented by dark pixels \cite{bay2013}. Therefore especially when the oil spill is close to land in an image, the correlation coefficient change statistics behaves poorly. Using a threshold of $T=0.6$, the change maps can be obtained with the probabilities of $P_d \approx 0.90$ and $P_{fa} \approx 0.35$ which can be read from the ROC curves discussed in the previous section. This high value of probability of false alarm is selected as a trade off of having a high probability of detection to capture the oil spill changes in the vicinity of the peninsula. However the correlation coefficient change statistic algorithm behaves poorly in detecting the changes due to oil spill despite this high value of the $P_{d}$ \cite{bay2013}. This poor performance can be seen in the Figures~\ref{CMGulf1} and~\ref{CMGulf12}. 

For the intensity ratio change statistics algorithm, the representative changed and unchanged areas lead to $R_0 \approx 0$ dB and  $R_1 \approx 5.05$ dB. The change maps can be obtained with the probabilities of $P_d \approx 0.95$ and $P_{fa} \approx 0.01$ using a threshold of $T=0.5$, which can be read from the ROC curves. The resulting change maps can be seen in the Figures~\ref{CMGulf1} and~\ref{CMGulf12} below. The intensity ratio change statistic performs excellent for the oil spill detection. This is due to the higher sensitivity of the intensity ratio measure to the changes in the dark pixels compared to the changes in the bright pixels. 

Additionally as it can be realized from the Figures~\ref{CMGulf1} and~\ref{CMGulf12} above and below, two final change map method proposed can be used as a double-check mechanism which reduces the probability of false alarm. This method works well for both of the algorithms used \cite{bay2013}. 

For the correlation coefficient change statistic the probabilities of false alarm and detection of the joint change map can be calculated as $P_{fa} \approx 0.35^2 = 0.12$ and $P_{d} \approx 0.90^2 = 0.81$ . For the intensity ratio change statistic the joint change map has  $P_{fa} \approx 0.01^2 = 0.0001$ and $P_{d} \approx 0.95^2 = 0.90$. Although the double change map methodology causes a small reduction in $P_{d}$, it reduces the $P_{fa}$ significantly therefore enhances the change detection \cite{bay2013}. For higher values of $P_{d}$ used in obtaining the two change maps, the joint change map will reduce $P_{d}$ insignificantly \cite{bay2013}.
 
\begin{figure}[h]
\begin{center}
   \includegraphics[width=3.4in]{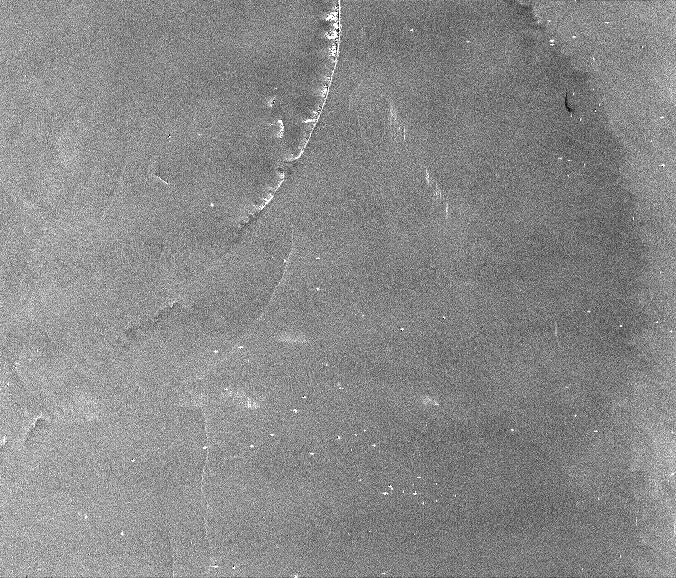}
\end{center}
\caption{Image ALPSRP193440580-Acquisition Date 09/11/2009}
\label{image_ALPSRP193440580}
\end{figure}

\begin{figure}[h!]
\begin{center}
   \includegraphics[width=3.4in]{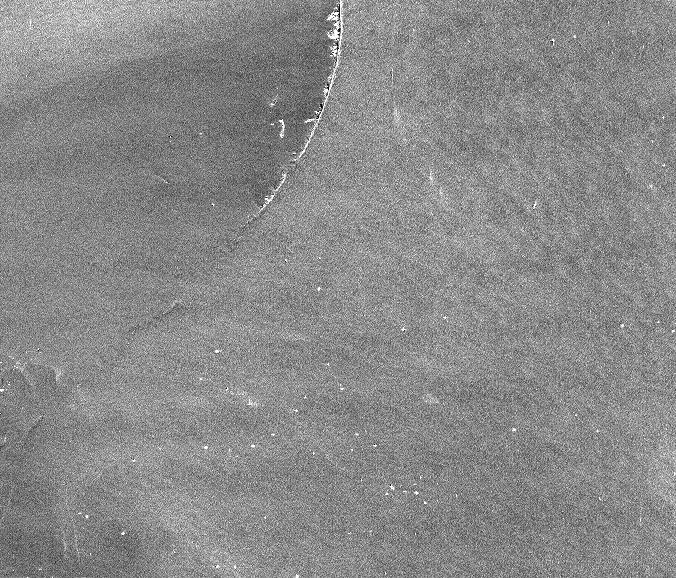}
\end{center}
\caption{Image ALPSRP200150580-Acquisition Date 10/27/2009.}
\label{image_ALPSRP200150580}
\end{figure}

\begin{figure}[h!]
\begin{center}
   \includegraphics[width=3.4in]{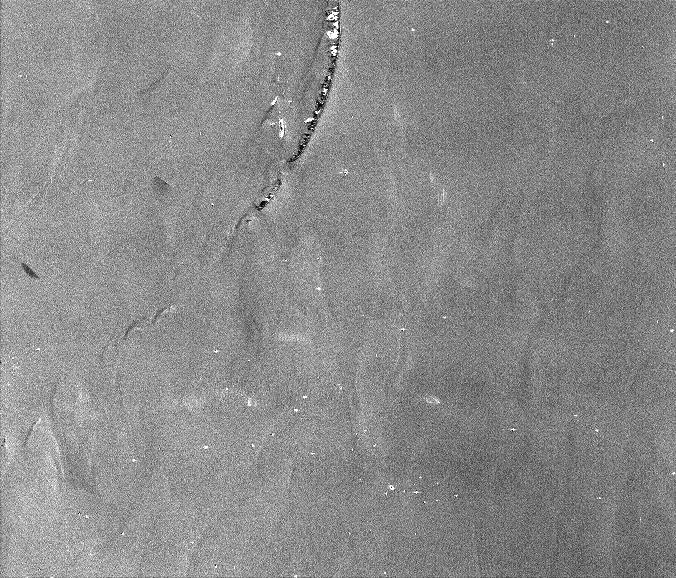}
\end{center}
\caption{Image ALPSRP240410580-Acquisition Date 07/30/2010.}
\label{image_ALPSRP240410580}
\end{figure}

\begin{figure}[h!]
\begin{center}
   \includegraphics[width=3.4in]{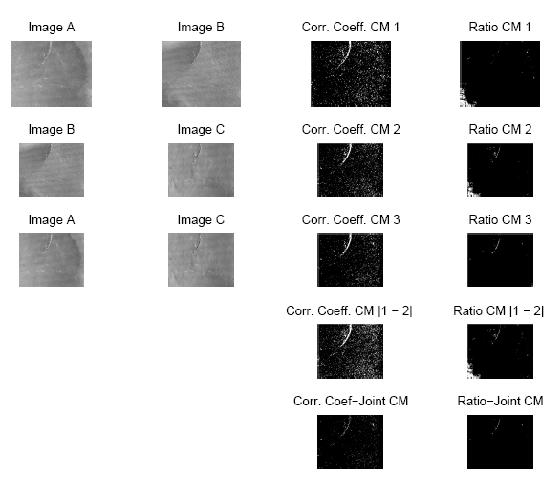}
\end{center}
\caption{Change maps of oil spill imagery collected on location 2.}
\label{CMGulf2}
\end{figure}

\begin{figure}[h!]
\begin{center}
   \includegraphics[width=3.4in]{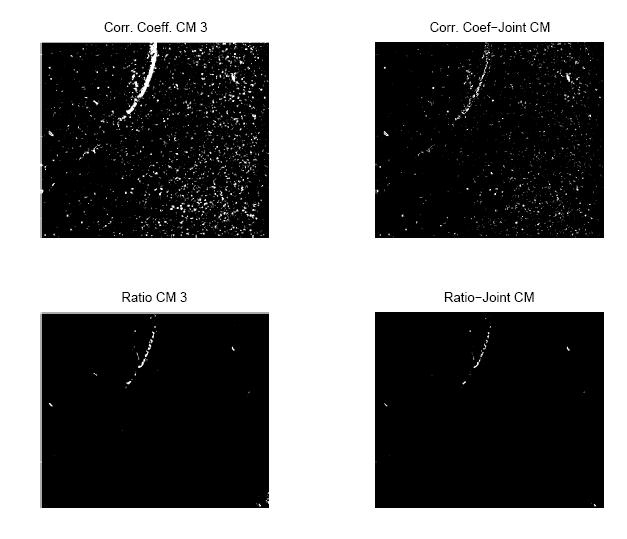}
\end{center}
\caption{Final change maps of oil spill imagery collected on location 2.}
\label{CMGulf22}
\end{figure}

The co-registered images of the Chandeleur Islands acquired at location 2 can be seen in the Figures~\ref{image_ALPSRP193440580},~\ref{image_ALPSRP200150580} and~\ref{image_ALPSRP240410580}. Applying the change detection process on this set of imagery, change maps in the Figures~\ref{CMGulf2} and~\ref{CMGulf22} can be obtained.



As it can be seen in change maps presented in the Figures~\ref{CMGulf2} and~\ref{CMGulf22}, the correlation coefficient change statistics algorithm for the imagery acquired at location 2 is again more sensitive to the changes on the bright areas due to land of Chandeleur Islands and therefore behaves poorly in oil spill detection. The intensity ratio change statistic is again very successful in capturing the oil spill changes which can be seen in the Figures~\ref{CMGulf2} and~\ref{CMGulf22} \cite{bay2013}.  

Additionally it can be realized from the figures that two final change map method can effectively be used to reduce the probability of false alarm for both of the algorithms.  

The probabilities of $P_{fa} \approx 0.35^2 = 0.12$ and $P_{d} \approx 0.90^2 = 0.81$ are the probabilities of the joint change map obtained by the correlation coefficient change statistic. For the intensity ratio change statistic these probabilities for the joint change map can be calculated as  $P_{fa} \approx 0.01^2 = 0.0001$ and $P_{d} \approx 0.95^2 = 0.90$. Although the joint change maps obtained exhibits a small amount of reduction in $P_{d}$ it can be very beneficial for reducing the $P_{fa}$ significantly. When higher values of $P_{d}$ is used in obtaining the two change maps, the joint change map will suffer less from the loss of $P_{d}$, almost to an insignificant level \cite{bay2013}.

  \begin{figure}[h!]
\begin{center}
   \includegraphics[width=3.4in]{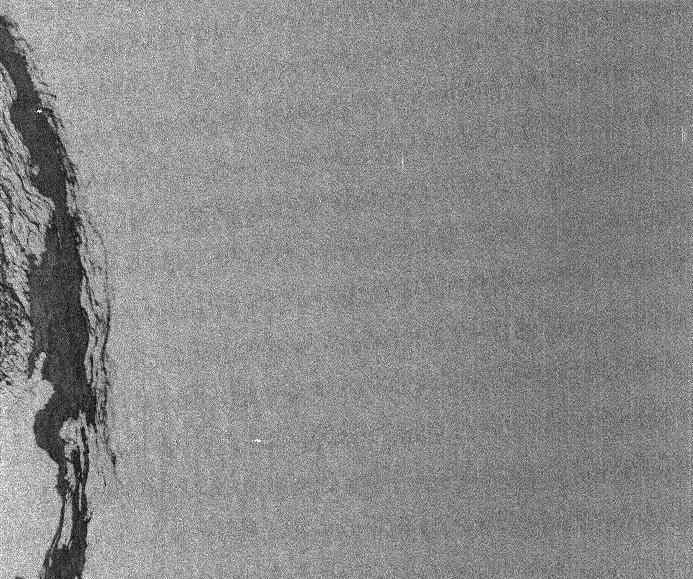}
\end{center}
\caption{Image ALPSRP231220550-Acquisition Date 05/28/2010.}
\label{image_ALPSRP231220550}
\end{figure}

 \begin{figure}[h!]
\begin{center}
   \includegraphics[width=3.4in]{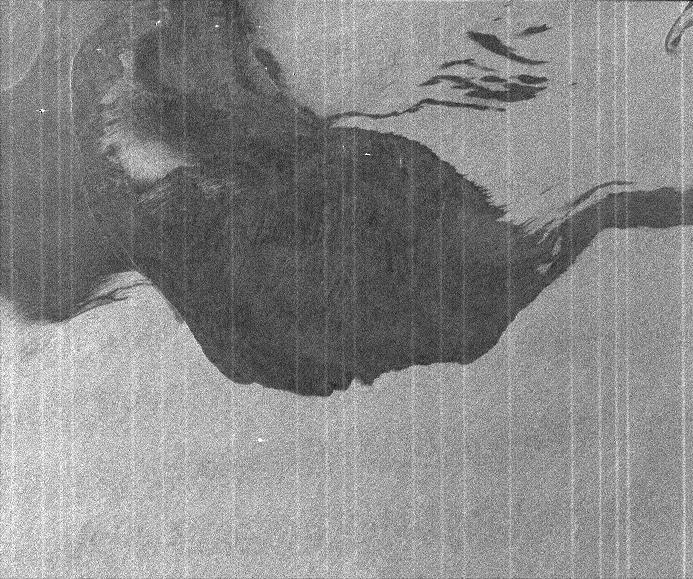}
\end{center}
\caption{Image ALPSRP237930550-Acquisition Date 07/13/2010.}
\label{image_ALPSRP237930550}
\end{figure}
	
	\begin{figure}[h!]
   \begin{center}
   \includegraphics[width=3.4in]{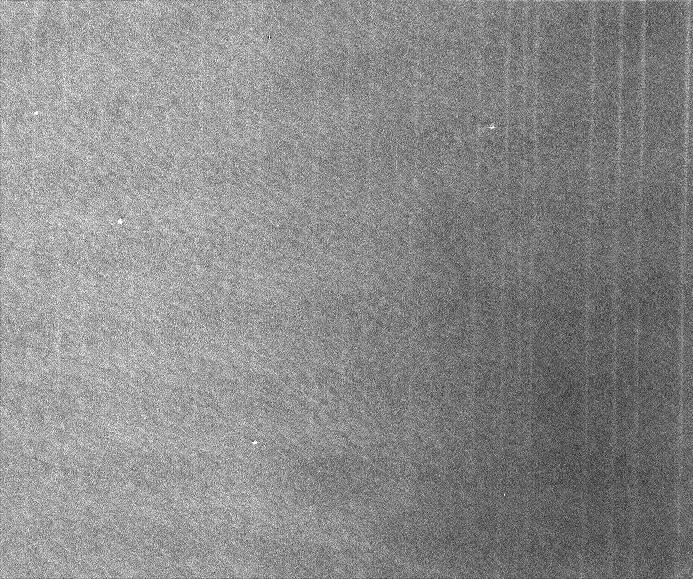}
\end{center}
\caption{Image ALPSRP244640550-Acquisition Date 08/28/2010.}
\label{image_ALPSRP244640550}
\end{figure}

It is useful to remember that Chandeleur Islands is a very dynamic coastal region which is subject to many coastal processes such as hurricanes, floods, waves and coastal erosion. Therefore they are subjected to significant erosion losses. It is useful to note that correlation coefficient change statistic can be used as a critical change measure for monitoring the coastal erosion amounts and rates \cite{bay2013}. 

In the Figures~\ref{image_ALPSRP231220550},~\ref{image_ALPSRP237930550} and~\ref{image_ALPSRP244640550}, the co-registered images acquired at location 3 can be seen.

For the correlation coefficient change statistics algorithm on this image set, the selection of representative changed and unchanged areas lead to $\gamma_{changed} \approx 0.30$ and  $\gamma_{unchanged}=0.69$, respectively. For a threshold of $T=0.5$, one can obtain the change maps with $P_d \approx 0.93$ and $P_{fa} \approx 0.08$. The parameters for the intensity ratio change statistics is the same as before \cite{bay2013}.
 
\begin{figure}[h]
   \begin{center}
   \includegraphics[width=3.4in]{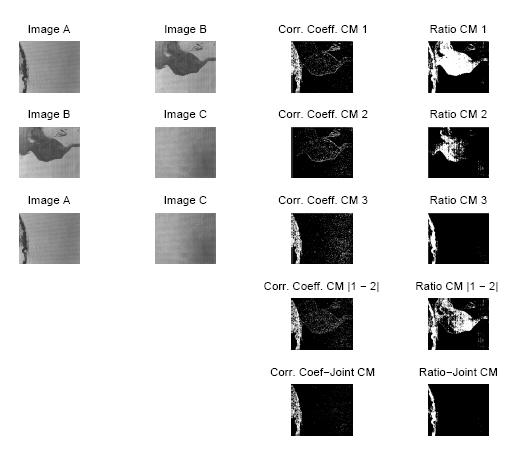}
\end{center}
\caption{Change maps of oil spill imagery collected on location 3.}
\label{CMGulf3}
\end{figure}

\begin{figure}[h]
   \begin{center}
   \includegraphics[width=3.4in]{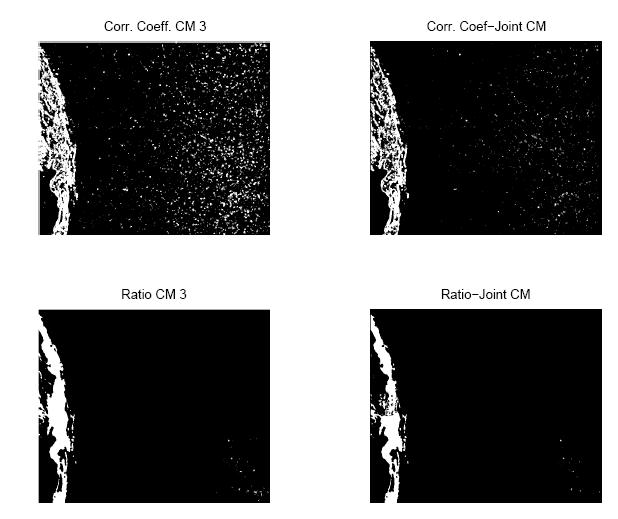}
\end{center}
\caption{Final change maps of oil spill imagery collected on location 3.}
\label{CMGulf32}
\end{figure}

As it can be seen in change maps presented in the Figures~\ref{CMGulf3} and~\ref{CMGulf32}, the intensity ratio change statistics still performs better. However the correlation coefficient change statistics is also successful in detecting the contaminated areas due to the oil spill. This is mainly because there is no land in the scene so that there are only very few bright pixels. For this reason, the correlation coefficient change statistics performs better compared to its previous performance on the previous image sets discussed above. Also the oil surfactant in the first image in the sequence is denser compared to the second image. This causes the CD methods to behave better since change measures get more different from the threshold values \cite{bay2013}. 

Additionally, again the two final change map method reduces the number of false alarms. This can be realized by comparing the change map between the first and the third images (CM 3) and the joint change map (Joint CM) obtained by intersecting the CM 3 and the cumulative additions of CM 1 and CM 2, both for the correlation coefficient and the intensity ratio change statistics.

For the correlation coefficient change statistic the probabilities of false alarm and detection of the joint change map can be calculated as $P_{fa} \approx 0.08^2 = 0.0064$ and $P_{d} \approx 0.93^2 = 0.87$ . For the intensity ratio change statistic, the probabilities for the joint change map can be calculated as  $P_{fa} \approx 0.01^2 = 0.0001$ and $P_{d} \approx 0.95^2 = 0.90$. Although the double change map method causes a small amount of reduction in $P_{d}$, it ia very beneficial for reducing the $P_{fa}$ significantly. When higher values of $P_{d}$ is used the joint change map will not suffer from the loss of $P_{d}$ \cite{bay2013}.

\section{Conclusion and Future Work}

In this study an assessment of the existing non-coherent change detection techniques used in remote sensing is performed for the oil spill detection. Two algorithms, the correlation coefficient change statistic and the intensity ratio change statistic algorithms are utilized for oil spill detection. These two algorithms are applied on SAR imagery of the 2010 Gulf of Mexico oil spill which are acquired by Japanese Aerospace Agency's Advanced Land Observing Satellite (ALOS). Additionally a new methodology is tested which can reduce the probability of false alarm. This methodology depends on the idea of generating two different final change maps of the initial and final images. First final change map is generated by cumulatively adding the sequences of change maps in a manner that common change areas are excluded and uncommon change areas are included. The second final change map is generated by comparing the first and the last images in the temporal sequence by the algorithm in use. Although this method can be generalized to longer temporal image sequences, in this paper temporal sequences with three imagery are used. This approach provides a double check mechanism to the conventional approach and thus reduces the probability of false alarm. 

It is shown that the intensity ratio change statistic performs significantly better compared to the correlation coefficient change statistic for detection of the oil spill, especially if the oil spill is in the vicinity of the land. Additionally it is shown that by taking the joint change map of the two final change maps, the double change map methodology reduces probability of false alarm significantly.

As a future work, other algorithms widely used in the literature can be tested for the detection of the oil spill in 2D or higher dimensions. The two final change map method employed in this paper can be tested with other change detection algorithms, except with those trivial ones which give the same result of the direct comparison of the first and the last image in the sequence, such as simple differencing or log-ratioing. Also testing this methodology with other CCD algorithms for oil spills and other type of disasters is a future research direction.

\section*{Acknowledgments}
The first author thanks NASA and JAXA for providing the satellite imagery during his Ph.D. studies at the Georgia Institute of Technology and thanks for the support of the I\c{s}\i k University.

\end{document}